\documentclass[useAMS,usenatbib]{mn2e}
\usepackage{natbib}
\usepackage{graphicx}
\usepackage{caption}
\usepackage{float}

\newcommand{\msun}{\mbox{$\,{\rm M}_\odot$}}

\title[The time-dependent galactic potential]{Galactic orbital motions of star clusters: static versus semicosmological time-dependent Galactic potentials}

\author[Haghi et al.]
{Hosein Haghi$^{1}$\thanks{
E-mail:  \mbox{haghi@iasbs.ac.ir} (HH)
\mbox{a.hasani@iasbs.ac.ir} (AHZ);
\mbox{saeed.taghavi.v@gmail.com} (ST)
 }, Akram Hasani Zonoozi$^{1}$, Saeed Taghavi$^{1,2}$\\\\
$^{1}$Department of Physics, Institute for Advanced Studies in Basic Sciences (IASBS), Zanjan 45137-66731, Iran\\
$^{2}$Department of Mathematics, Mazandaran University of Science and Technology, Behshahr 48518-78413, Iran\\
}

\begin{document}

\date{Accepted ....; Received ...}
%\date{Accepted 2014 August 20. Received 2014 August 20; in original form 2014 June 5}

\pagerange{\pageref{firstpage}--\pageref{lastpage}} \pubyear{2013}

\maketitle

\label{firstpage}

\maketitle

\begin{abstract}

In order to understand the orbital history of Galactic halo objects, such as globular clusters, authors usually assume a static potential for our Galaxy with parameters that appear at the present-day. According to the standard paradigm of galaxy formation, galaxies grow through a continuous accretion of fresh gas and a hierarchical merging with smaller galaxies from high redshift to the present day. This implies that the mass and size of disc, bulge, and halo change with time. We investigate the effect of assuming a live Galactic potential on the orbital history of halo objects and its consequences on their internal evolution. We numerically integrate backwards the equations of motion of different test objects located in different Galactocentric distances in both static and time-dependent Galactic potentials in order to see if it is possible to discriminate between them. We show that in a live potential, the birth of the objects, 13\,Gyr ago, would have occurred at significantly larger Galactocentric distances, compared to the objects orbiting in a static potential.  Based on the direct $N$-body calculations of star clusters carried out with collisional $N$-body code, \textsc{nbody6}, we also discuss the consequences of the time-dependence of a Galactic potential on the early- and long-term evolution of star clusters in a simple way, by comparing the evolution of two star clusters embedded in galactic models, which represent the galaxy at present and 12\,Gyr ago, respectively. We show that assuming a static potential over a Hubble time for our Galaxy as it is often done, leads to an enhancement of mass-loss, an overestimation of the dissolution rates of globular clusters, an underestimation of the final size of star clusters, and a shallower stellar mass function.

\end{abstract}


\section{Introduction}\label{Sec:Intro}

Tens of satellite galaxies and about 160 globular clusters (GCs; \citealt{Harris96, Harris10}), have been identified in the Milky Way (MW), that are distributed out to more than 200 kpc, orbiting around the centre of our Galaxy. Observations show that nearly all galaxies host these systems, with giant ellipticals having almost the largest population \citep{Brodie06}.  Stellar population studies have revealed that GCs have ages up to 13\,Gyr (e.g., \citealt{Chaboyer02, Hansen02}), and therefore they represent fossil records of the earliest epoch of galaxy formation.  As such, they are potentially powerful probes of physical conditions in the high redshift Universe \citep{Brodie06}.

All star clusters lose mass over time and this depends on a number of internal and external processes, as e.g.,  mass-loss due to stellar evolution, mass segregation, and core collapse due to two-body relaxation, and the external tidal field of the parent galaxy within which the cluster orbits \citep{Vesperini97, Baumgardt03, Heggie03, Gieles11}. In addition, the evolution of star clusters depends crucially on the initial conditions (e.g., the initial mass profile of the star cluster, the initial mass function (IMF) of the stars, and the initial binary fraction) and the orbital parameters of the cluster \citep{ Madrid12, Webb13, Webb14, Haghi14}.

Based on $N$-body simulations of GC systems, it is well accepted that the GC populations we observe today are only the very remnants of much richer systems (e.g.,~\citealt{Bonaca12, Grillmair13, Brockamp14, Koposov14}). The rate of GC erosion strongly depends on the details of the gravitational potential of the host galaxy as well as on the internal properties of the GCs \citep{Brockamp14}. Therefore, the present-day distribution of GC systems around the MW, and their properties may be valuable probes of the Galaxy potential.

The survival or dissolution of star clusters in the galactic tides within which they orbit also depends crucially on their orbital history: star clusters with large radii spending the major part of their lifetime in the innermost regions of our Galaxy are more susceptible to tidally induced mass-loss, whereas the outer halo objects can survive for a Hubble time (e.g., \citealt{Giersz97, Hurley07,  Heggie08, Gieles11, Brockamp14, Haghi14}). Also, calculation of orbital motions backward in time is necessary to model the formation of the stellar and gaseous streams emerging from GCs or accreting satellite galaxies.

Therefore, a detailed understanding of the orbital history which requires a better understanding of the evolution of the Galactic potential since its formation, is an essential issue in investigating the tidal erosion of GCs (via mass-loss) and accreting disruption of satellite galaxies. Many authors usually use the static potential, i.e.,  assume that it remains unchanged during the orbital integration. But, observations have revealed that the size and the mass content of galaxies change significantly with redshift such that the sizes of the galaxies at high redshifts are smaller in comparison with galaxies of similar mass in the local universe (e.g.,\citealt{Franx08, Williams10, Mosleh11, Law12, Mosleh13}).

There are many proposed scenarios to explain the physical processes of galaxy assembly that well reproduce the observable properties like, e.g., the stellar mass and size of galaxies at different redshifts. Among them are the galaxy minor or major mergers (e.g., \citealt{Khochfar06, Khochfar09, Naab09}), and the accretion of fresh gas in outer regions activating new star formation (e.g., \citealt{Elmegreen08, Dekel09}). Indeed, in the standard picture of galaxy formation, galaxies are embedded in massive virialized haloes of dark matter \citep{Springel06}. These dark matter haloes accumulate over time hierarchically, continuously growing via accretion of dark matter and merging with other haloes from high redshift to the present day. The fraction of GCs that have survived to the present day has evolved in a time-dependent potential of the host galaxy.

Understanding the influence of the time-dependence of the Galactic potential on the orbital history of the halo objects (e.g., GCs) rotating around the MW at different Galactocentric distances, and its consequence on their early- and long-term evolution over a Hubble time is the main motivation for this paper.

Indeed, because of the wide range of time-scales and size scales, from the two-body strong encounters of stars to the galactic scales, the $N$-body simulations of star clusters in a live galactic potential are challenged. However, attempts have been made to overcome this difficulty.  A pioneering study of this subject was carried out by \cite{Renaud11} who investigated the evolution of star clusters including a time-dependent potential. They proposed a novel approach to extract the tidal information as tables of tensors from a galaxy or cosmology simulation along one orbit. Their method has been applied to a large number of star clusters in a galaxy major merger \citep{Renaud13} emulating the Antenna galaxies (NBC 4038/39). A more recent improvement of this method using any definition of the external potential as a function of space and time can be found in \cite{Renaud15}.  Similar work was done by \cite{Rieder13} who tracked the tidal history of clusters in a cosmological context by inserting clusters into a dark matter only cold dark matter simulation. They also found that mergers tend to increase the mass-loss rates of clusters.

In this work, we will estimate how the mass, characteristic radii, and the mass function (MF) slope of a star cluster would change if it is evolved in the time-dependent potential by comparing the evolution of two star clusters embedded in two galactic models, which represent our Galaxy with the present-day parameters and with parameter values at 12\,Gyr ago, respectively.

As a concrete example of motion within the Galactic halo, we will also use the backward motion of the Large Magellanic Cloud (LMC) located at about 50 kpc from the centre of the MW in both static and live Galactic potentials to see if, at least in principle, our approach is able  to discriminate  between them.

In Section 2, we describe the characteristics of the time-dependent galactic potential we have used in this paper.  We compare the orbital motions of different test particles around the centre of the Galaxy in Section 3. In Section 4, we present results of $N$-body calculations for the dynamical evolution of star clusters moving through an external galaxy with different background potential parameters. The simulations were carried out with the collisional N-body code \textsc{nbody6} on desktop workstations with Nvidia 690 Graphics Processing Units at the Institute for Advanced Studies in Basic Sciences (IASBS). Finally, in Section 5, we summarize our results.

\section{The Galactic potential} \label{potential}

\subsection{Static Galactic potential}

\begin{figure*}
\includegraphics[width=85mm]{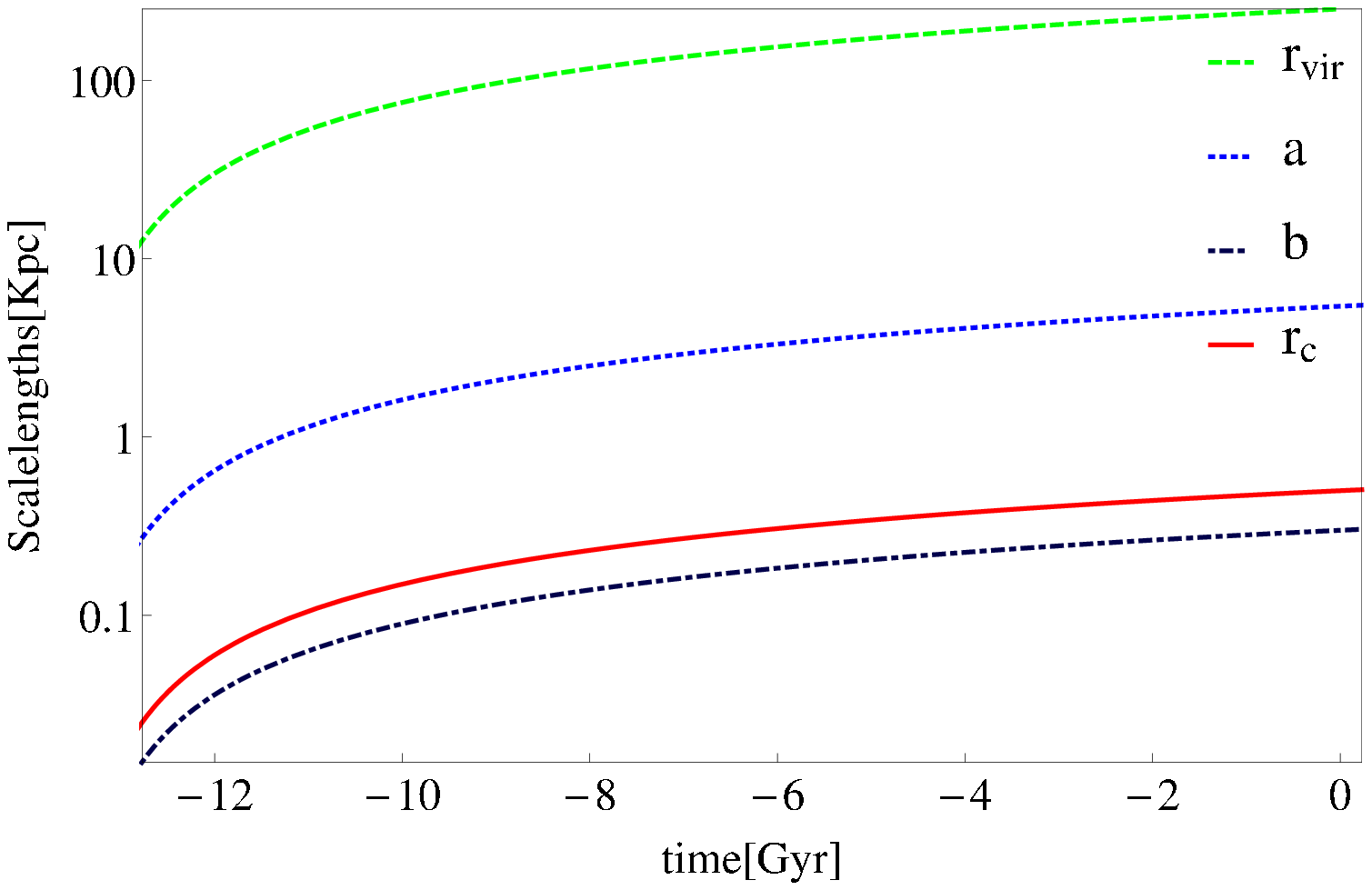}
\includegraphics[width=85mm]{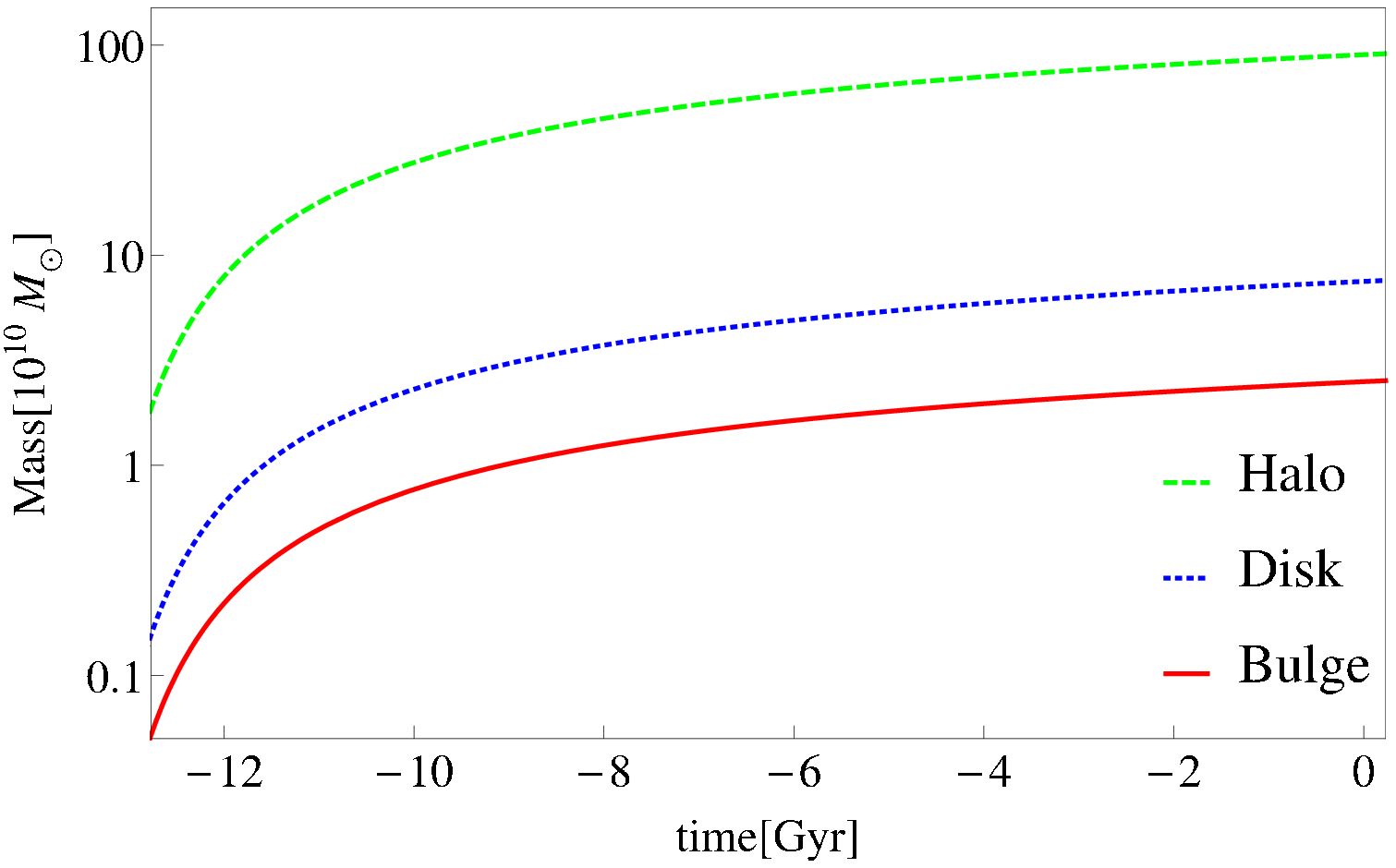}
\caption{Left-hand panel: logarithmic plot of the size scale of the MW's mass components as a function of cosmological time. The virial radius of the dark matter halo $r_{vir}$ (dashed green), scale radius of the disc $a$ (dotted blue), the scale-height of the disc $b$ (dash-dotted dark blue), and the scale radius of the bulge $r_{c}$ (solid red) are shown in the figure.  Right-hand Panel: the evolution of the mass scale of the MW's mass components as a function of cosmological time. The virial mass of the dark matter halo $M_{vir}$ (dashed green), the scale mass of the disc $M_d$ (dotted blue), and of the bulge $M_{b}$ (solid red) are shown in the figure. All components show fast growth in the past time.}
\label{fig:sizescale}
\end{figure*}

In the first part, for the Galaxy potential we, like many other authors, assumed that the galaxy potential is static and consists of three idealized components, %To model the Milky Way potential we chose a three-component system
\begin{equation}
	\Phi_{tot}=\Phi_{d}+\Phi_{b}+\Phi_{h},
\end{equation}
 including a Miamoto--Nagai disc potential \citep{miyamoto75} given by
\begin{equation}
\Phi_d (x,y,z) = -\frac{G M_d}{\sqrt{x^2+y^2+\left(a + \sqrt{z^2+b^2}\right)^2}},
\end{equation}
a central bulge, the Hernquist (1990) model,  given by
\begin{equation}
	\Phi_b (x,y,z) = -\frac{G M_b}{r+r_c},
\end{equation}
and a NFW dark matter halo \citep{NFW97} with a potential of the form
\begin{equation}
	\Phi_{NFW}=-\frac{G M_{vir}}{r [\log (1+c)-\frac{c}{1+c}]} \log \left( 1+\frac{c \ r}{r_{vir}} \right).
\end{equation}
Here, $r=\sqrt{x^2+y^2+z^2}$ is the distance from the galactic centre at any given time; $M_b$, $M_d$ and $M_{vir}$ are the characteristic mass of the bulge, disc, and halo, respectively; $r_c$ and $r_{vir}$ are the characteristic radius of the bulge and halo, respectively; and $a$ is the scale radius and $b$ the scale height that adjust the shape of the disc. The present-day numerical values of these parameters at redshift $z=0$ are given in the Table \ref{table:parameters}.

%\citep{sofue09}.
\begin{table}
	% title of Table
	\centering % used for centering table
	\begin{tabular}{c c c } % centered columns (4 columns)
		\hline\hline %inserts double horizontal lines
		Disc & Bulge & Halo  \\ [0.5ex] % inserts table
		%heading
		\hline % inserts single horizontal line
		$M_{d}=7.5 \times 10^{10}$ & $M_{b}=2.5 \times 10 ^{10} $ & $M_{vir}=9 \times 10^{11}$ \\ % inserting body of the table
		$a=5.4; \ b=0.3$ & $r_{c}=0.5$& $r_{vir}=250$ \\
		&  & $c=13.1$\\[1ex] % [1ex] adds vertical space
		\hline %inserts single line
	\end{tabular}
	\caption{The present-day parameters of the mass components of  the MW-like potential used in our calculations. Masses and distances are in M$_{\odot}$  and in kpc, respectively.}
		\label{table:parameters}
\end{table}

\begin{table}
	% title of Table
	\centering % used for centering table
	\begin{tabular}{c c c } % centered columns (4 columns)
		\hline\hline %inserts double horizontal lines
		Disc & Bulge & Halo  \\ [0.5ex] % inserts table
		%heading
		\hline % inserts single horizontal line
		$M_{d}=6.6 \times 10^{9}$ & $M_{b}=2.2 \times 10 ^{9} $ & $M_{vir}=7.9 \times 10^{10}$ \\ % inserting body of the table
		$a=0.65; \ b=0.04$ & $r_{c}=0.06$& $r_{vir}=29.9$ \\
		&  & $c=2.86$\\[1ex] % [1ex] adds vertical space
		\hline %inserts single line
	\end{tabular}
	\caption{The parameters of the mass components of the MW-like potential at $T=-12$\,Gyr. Masses and distances are in M$_{\odot}$  and in kpc, respectively. Eqs. 5 - 9 are used to calculate these values at $z=3.5$ (i.e., $T=-12$\,Gyr).}
		\label{table:parameters-12Gyr}
\end{table}

\subsection{Time-dependent Galactic potential}

To model the evolution of the Galactic potential, we assume a semicosmological time-dependent gravitational potential in which the characteristic parameters vary in time. The evolution of the mass and the virial concentration of the galaxy's halo as a function of redshift are given by \citep{Wechsler02,zhao03,Gomez10}
\begin{equation}
	M_{vir}(z)=M_{vir}(0) \exp(-2a_c z),
\end{equation}
where the formation epoch is set to be $a_c=0.34$, and
\begin{equation}
	c(z)=\frac{c(0)}{1+z}.
\end{equation}
For the disc and bulge we follow the recipe given by \cite{bullock05}, as for masses we have
\begin{equation}
	M_{d,b}(z)=M_{vir}(z) \frac{M_{d,b}(0)}{M_{vir}(0)},
\end{equation}
and for scale-lengths the evolution can be expressed as
\begin{equation}
	\{a,b,r_{c}\}(z)=r_{vir}(z) \frac{\{ a,b,r_{c} \}(0)}{r_{vir}(0)}.
\end{equation}
Here $r_{vir}$ is the virial radius of the dark matter halo, varying as
\begin{equation}
r_{vir}(z)=\left( \frac{3 M_{vir}(z)}{4\pi \Delta_{vir}(z) \rho_{c}(z)} \right)^{1/3},
\end{equation}
where $\Delta_{vir}(z)$ denotes the virial overdensity ,
\begin{equation}
\Delta_{vir}(z)=18\pi^{2}+82[\Omega(z)-1]-39[\Omega(z)-1]^{2}
\end{equation}
with $\Omega(z)$ the mass density of the universe,
\begin{equation}
\Omega(z)=\frac{\Omega_{m,0}(1+z)^{3}}{\Omega_{m,0}(1+z)^{3}+\Omega_{\Lambda,0}},
\end{equation}
and $\rho_{c}(z)$ is the critical density of the universe at a given redshift,
\begin{equation}
\rho_{c}(z)=\frac{3 H^{2}(z)}{8 \pi G}
\end{equation}
with
\begin{equation}
H(z)=H_{0}\sqrt{\Omega_{\Lambda,0}+\Omega_{m,0}(1+z)^{3}}.
\end{equation}
We also adopted a flat cosmology defined by $\Omega_{m,0}=0.3$ and $\Omega_{\Lambda,0}=0.7$ with a Hubble constant of $H(z=0)=H_{0}=70 \textnormal{kms}^{-1}\textnormal{Mpc}^{-1}$.\\

In cosmology one can label the time $t$ since the big bang in terms of the redshift of light emitted at $t$. By integrating the Friedmann equation, the behaviour of the cosmological redshift in terms of time for a flat universe can be found as
\begin{equation}
	z=\Bigg(\frac{\Omega_{m,0}\sinh^2(\frac{3}{2}H_0t\sqrt{\Omega_{\Lambda,0}})}{\Omega_{\Lambda,0}}\Bigg)^{-1/3}-1,
\end{equation}

In Fig. \ref{fig:sizescale}, we display how the characteristic parameters of the MW (i.e., all Galactic mass components and scalelengths) vary as a function of cosmological time. The parameters of the mass components of the MW-like potential at $T=-12$\,Gyr ($z=3.5$) are given in Table 2.  %Fig. \ref{tz} shows the evolution of the cosmological redshift as a function of time for a flat universe which is valid for all $z$.

\begin{figure}
	\includegraphics[width=85mm]{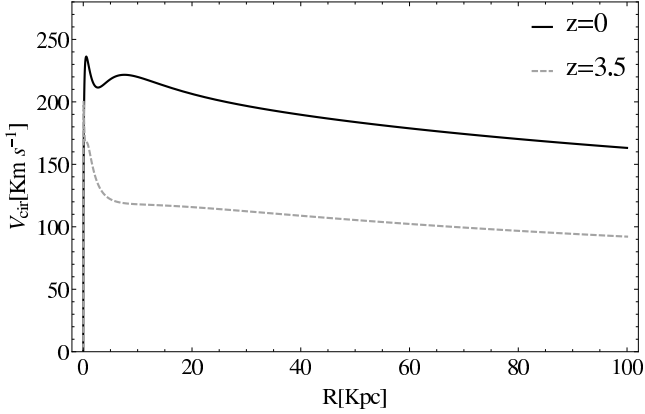}
	\caption{Comparison of the circular velocity curve as a function of Galactocentric distance  at $z=0$ (black solid line) with rotation curve at $z=3.5$ (grey dashed line). The asymptotic values of circular velocity in two frame work differ by about 75 kms$^{-1}$ after about 12\,Gyr of evolution.  }
		\label{fig:rotationcurve}
\end{figure}

The Galactic rotation curves deduced from the three-component mass model, described in Eqs. 2-13, at $z=0$ and $z=3.5$  are plotted in Fig. \ref{fig:rotationcurve}. As can be seen, the present-day circular velocity curve of this model takes a value of about 221 $\textnormal {kms}^{-1}$ at 8.5 $ \textnormal{kpc}$ from the galactic centre, while it is about  117 $\textnormal {kms}^{-1}$  at $z=3.5$. The smaller value of asymptotic rotational velocity at $R_G=100 $ kpc occurs in the time-dependent gravitational potential, while the static potential yields the larger one,  differing  by about  75 $\textnormal {kms}^{-1}$ after 12\,Gyr of evolution.

%\begin{figure}
%	\includegraphics[width=85mm]{tz_1.eps}
%	\caption{The cosmological redshift as a function of time for a WMAP flat universe. The redshift at $t=-12$ Gyr is about $z=3.5$. }
%		\label{tz}
%\end{figure}

\begin{figure}
	\includegraphics[width=80mm]{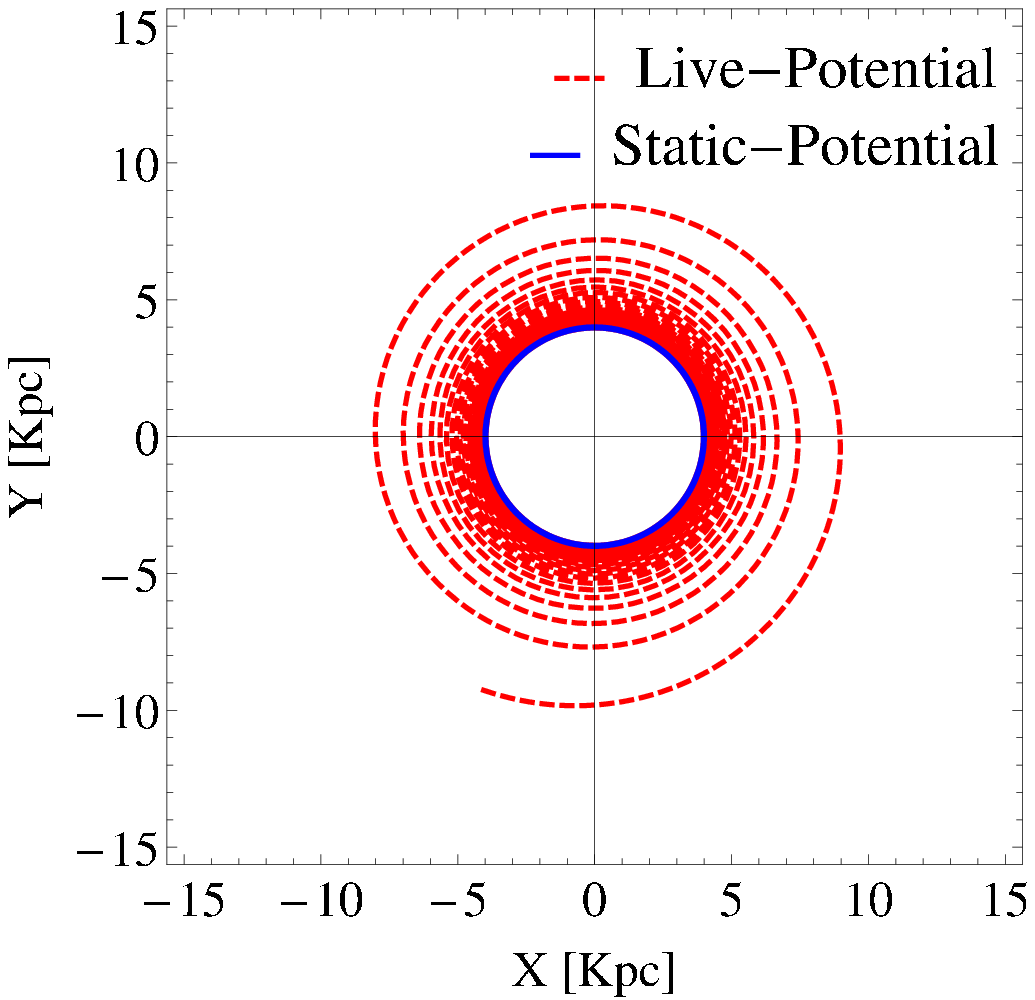}
    \includegraphics[width=80mm]{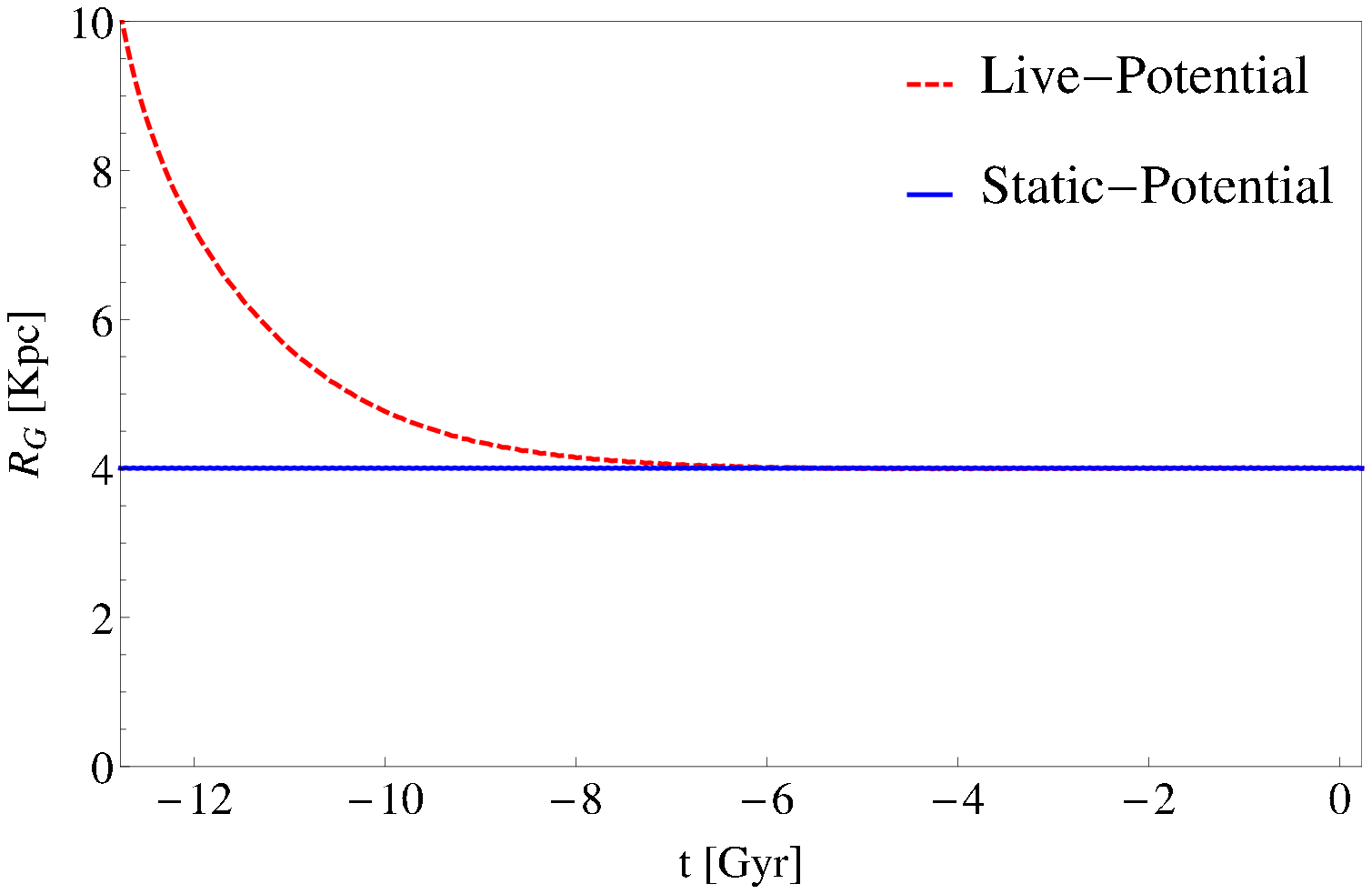}
	\caption{Upper panel: planar shape of the orbits of a test particle w.r.t. the MW, starting from $R_G=4$ kpc (the first line of Table \ref{table:virtualobjects_orbital_param}) . The red dashed line shows the trajectory of test object within the time-dependent potential while the blue line is for the static potential.  Lower panel: the time evolution (for the past 13 Gyr) of the radial distance to the Galactic centre of a test particle, starting from $R_G=4$ kpc  in the time-dependent potential (red dashed line) and the static potential (blue line). }
		\label{fig:YX_inner}
\end{figure}

\begin{figure}	
	\includegraphics[width=80mm]{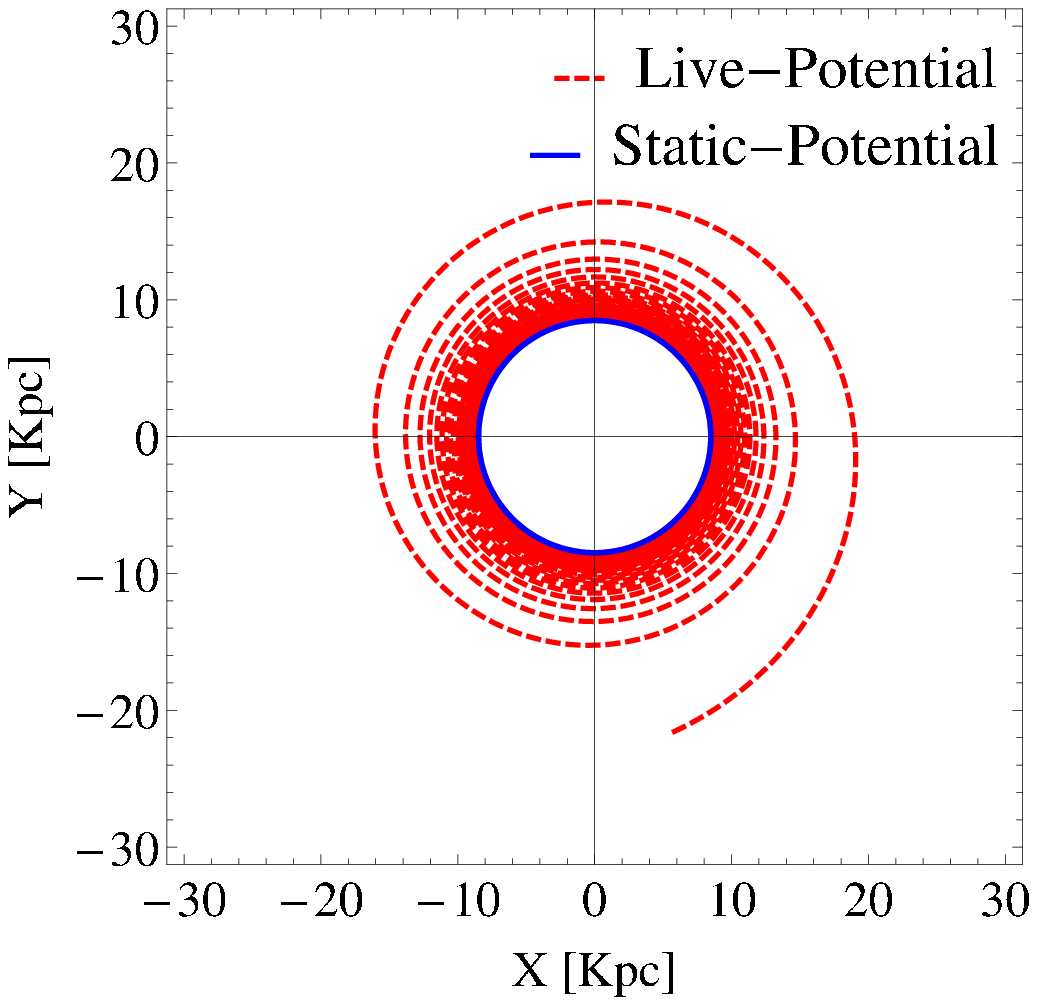}
    \includegraphics[width=80mm]{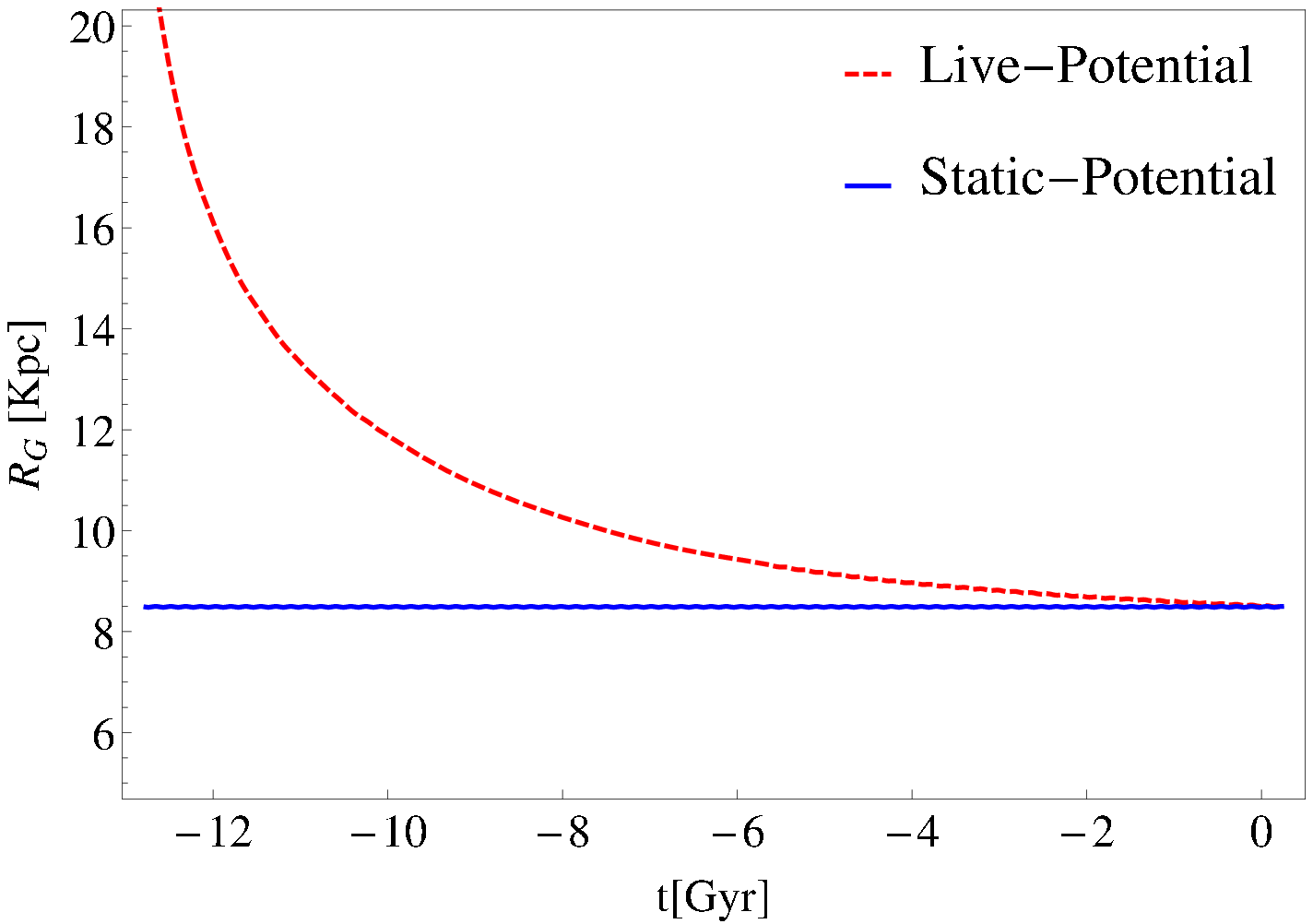}
	\caption{The same as Fig. \ref{fig:YX_inner} but for a test object that located initially at $R_G=8.5$ kpc as shown in the second line of Table \ref{table:virtualobjects_orbital_param}. }
	\label{fig:YX_Sun}
\end{figure}

\begin{figure}	
	\includegraphics[width=80mm]{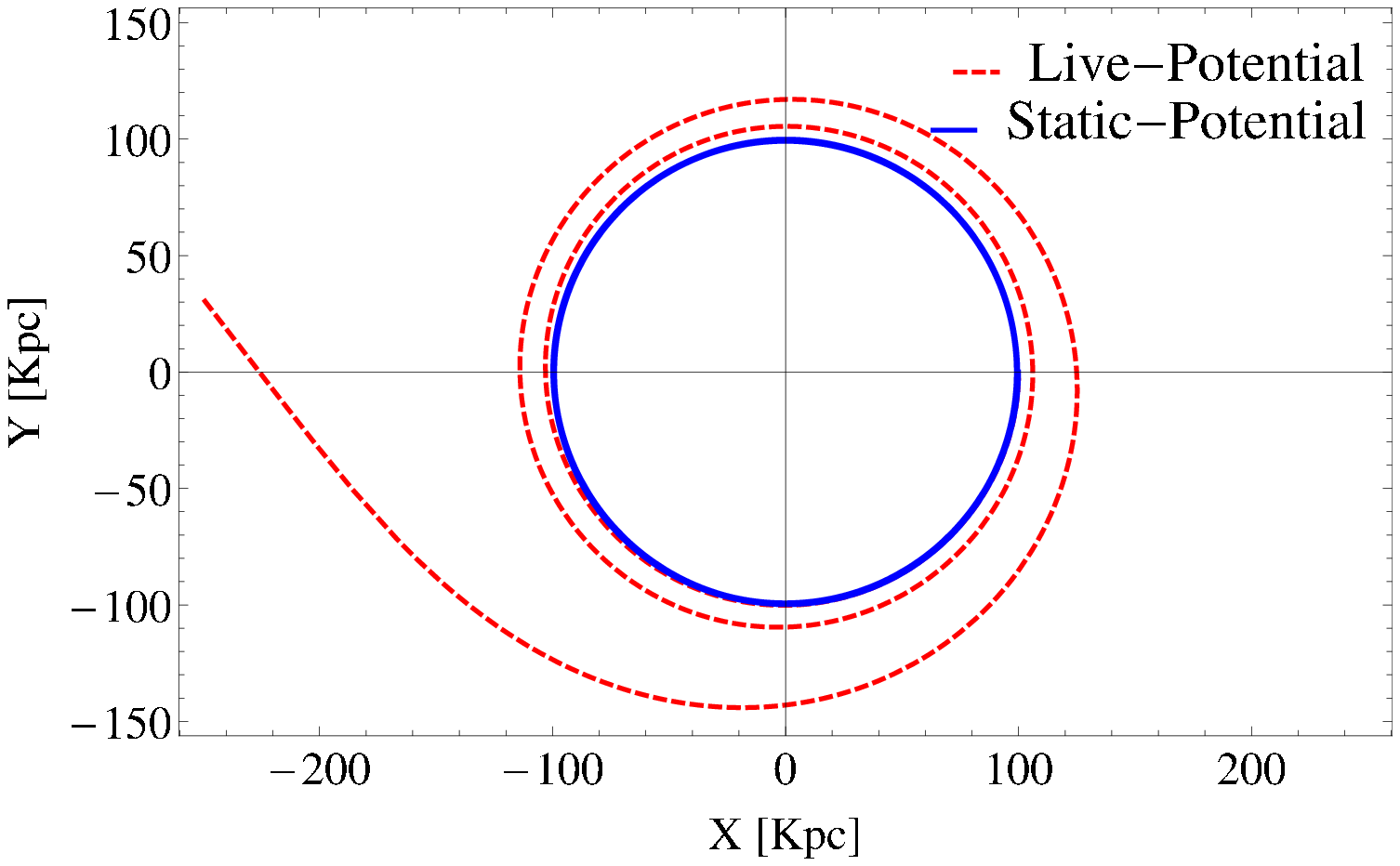}
    \includegraphics[width=80mm]{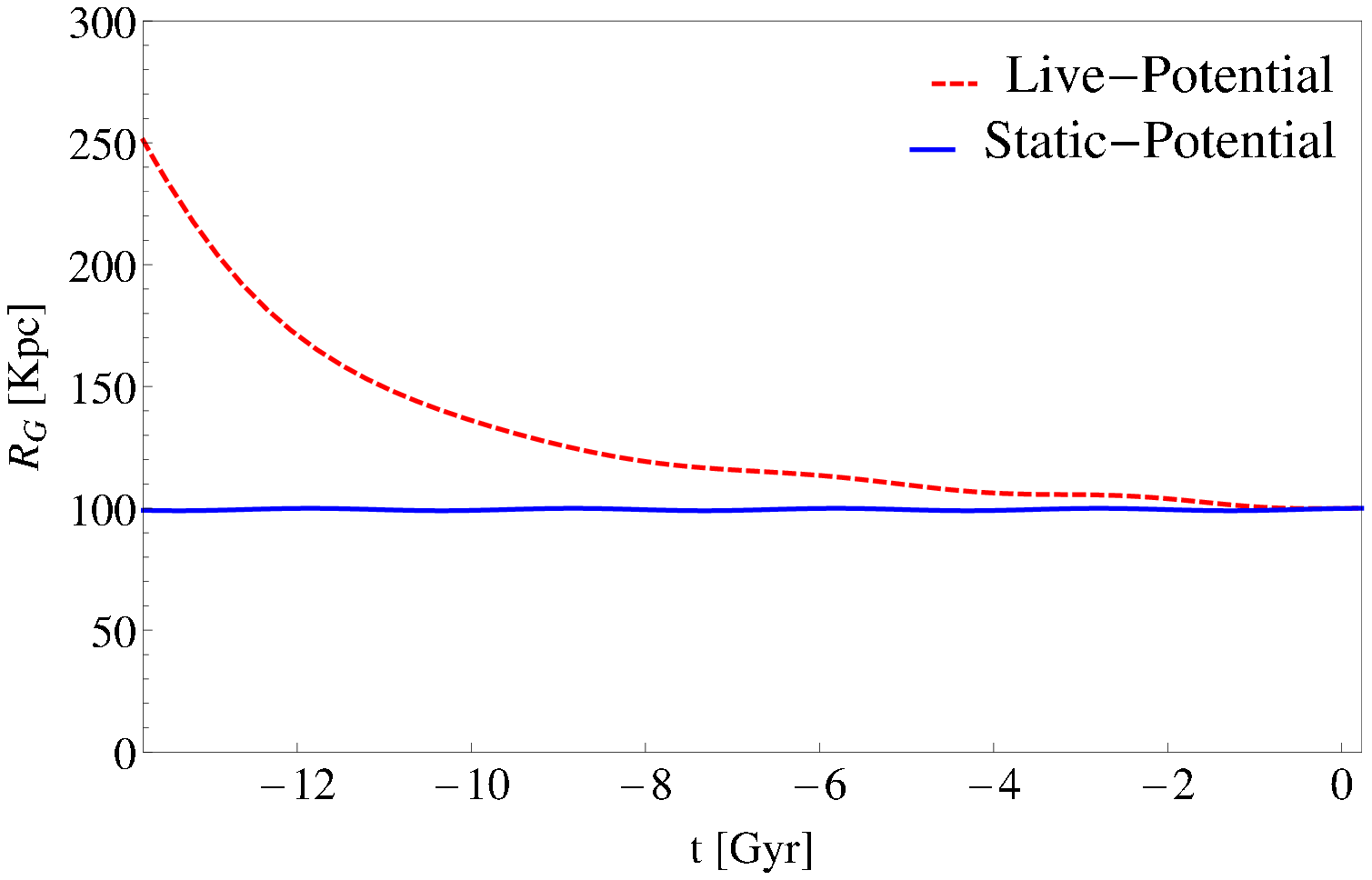}
	\caption{The same as Fig. \ref{fig:YX_inner}, but for a test object initially located at $R_G=$100 kpc as shown in the third line of Table \ref{table:virtualobjects_orbital_param}. }
	\label{fig:YX_outer}
\end{figure}
\begin{figure}
	\includegraphics[width=85mm]{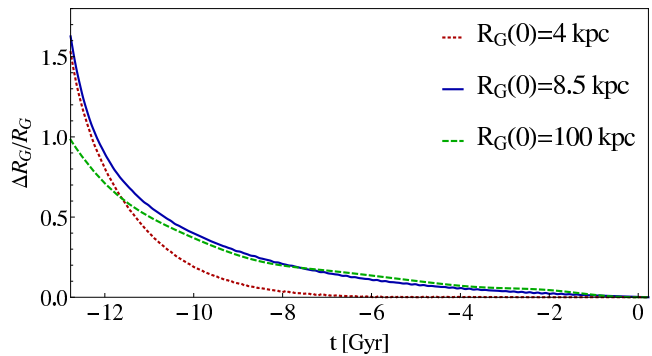}
	\caption{The evolution of the relative difference of the Galactocentric distance in the static and live Galactic potential in time that are plotted for individual test objects moving in different initial distances from the centre of Galaxy as given in Table \ref{table:virtualobjects_orbital_param}. The differences are small at the beginning of orbital backward motion and starts to increase at about $t=-6$ Gyr. The value of Galactocentric distance get doubled within a Hubble time for a test object initially located at  $R_G=$ 100 kpc, while it increases by a factor of about 2.5 for other two test objects at  $R_G=$ 4 and 8.5 kpc.}
	\label{fig:drt}
\end{figure}

\section{Motions backward in time}

In this section, we describe the results from the numerical integration of the equation of motion to find the trajectory of a test object located in different Galactocentric distances moving under  the static Galactic potential and compare them with the same calculations in a time-dependent Galactic potential by using  Eqs. 5 - 13 and the present-day parameters given in Table 1.

Using the gravitational potential components described above, we will numerically integrate the three scalar differential equations, written in Cartesian coordinates, corresponding to the vector differential equation,

\begin{equation}
	\ddot{\bf{r}}=-\bf{\nabla}\Phi_{tot}.
\end{equation}
Here we use the fourth-order Runge--Kutta method to calculate the equation of motion of test objects and then to extract the trajectory of object by backtracking orbit from its current position and velocity. Using the initial conditions of Table \ref{table:virtualobjects_orbital_param}, the equations of motion are then numerically integrated backward in time for 13\,Gyr in both live and static gravitational potentials. In this manner, we therefore obtain sets of initial positions and velocities required for forward integration in time and can be applied in e.g., $N$-body simulations of realistic GCs or satellite galaxies.

We calculate the orbit for three test objects:  an inner test object currently located  at $R_G=$ 4 kpc,  a solar-distance object located at $R_G=$8.5 kpc, and an outer object with the present-day Galactocentric distance of $R_G=$ 100 kpc. For all objects, we calculate the trajectory within two different Galactic models: the static Galactic potential and time-dependent Galactic potential. The orbital parameters of these test objects are summarized in Table \ref{table:virtualobjects_orbital_param}.

The initial velocities are extracted from the present-day rotation curve. That is, the orbits are circular in the static potential, while in the live potential the Galactocentric distances increase looking backwards in time. Note that, the test objects evolve on circular orbits at different galactocentric distances in the disc plane (i.e., the inclination angle of the orbits w.r.t the galactic disc is zero). In Figs. \ref{fig:YX_inner} - \ref{fig:YX_outer}, we plot the orbital sections in the Galactocentric coordinate planes of the numerically integrated trajectories  of the test objects from now to 13 Gyr ago. The time evolution of distances from the centre of the MW for all test objects are shown in the lower panel of  Figs. \ref{fig:YX_inner} - \ref{fig:YX_outer}.  As can be seen in  Fig. \ref{fig:drt}, the Galactocentric distances of these objects at $t=-13$ Gyr are larger than initial values at $t=0$ by a factor of about $~2$ (for test object initially located at  $R_G=$ 100 kpc), and 2.5 (for test objects initially located at 4 and 8.5 kpc). It should be noted such a large difference in determining the birth place of realistic objects (like e.g., GCs in the MW) may pose problems concerning the dynamical evolution. I will be back to this important issue in more details in Sec. \ref{dynamic}.

\begin{table}
	% title of Table
	\centering % used for centering table
	\caption{Initial coordinates, in kpc, and initial velocities, in $kms^{-1}$, in Galactocentric rest frame adopted for different test objects used in this paper. We have used the present-day rotation curve of our Galaxy for initial circular velocities. }
	\begin{tabular}{c c c } % centered columns (4 columns)
		\hline\hline %inserts double horizontal lines
		&$r(x,y,z)$ & 3D $v(x,y,z)$   \\ [0.5ex] % inserts table
		%heading
		\hline % inserts single horizontal line
		model1 (inner-part object)&(4,0,0)      & (0,215,0)\\
		model2 (solar-distance)&(8.5,0,0)      & (0,221,0)\\
        model3 (outer-part object)&(100,0,0)      & (0,195,0)\\
		\hline %inserts single line
	\end{tabular}
	\label{table:virtualobjects_orbital_param} % is used to refer this table in the text
\end{table}

\subsection{Motion of the LMC backward in time}

As a concrete example of motion within the Galactic halo, let us consider the orbital motion of the LMC currently located at about 50 kpc from the centre of the MW. Many authors have shown that the position of the Magellanic Stream (MS) \footnote{A narrow band of neutral hydrogen clouds lies along a
great circle from $ (l= 91^{\circ}; b=-40^{\circ}) $ to $ (l= 299^{\circ}; b=-70^{\circ})$, started from the MCs and oriented towards the South Galactic Pole.} follows the orbits of MCs ( see e.g. \citealt{Moore94, Connors06, Haghi06, Besla07, Haghi09, Haghi10}). Moreover,  the shape and the kinematics of the MS  is strongly influenced by the overall properties of the underlying potential \citep{Murai80, Lin82, Heller94, Sofue94, Gardiner96}. Therefore, it is most striking  to compare the trajectory of the LMC with orbits predicted in both static and live Galactic potentials. Table \ref{table:LMC_orbital_param} summarizes the present-day Galactocentric Cartesian coordinates and velocities of the LMC \citep{M05}.

\begin{figure}
	\includegraphics[width=80mm]{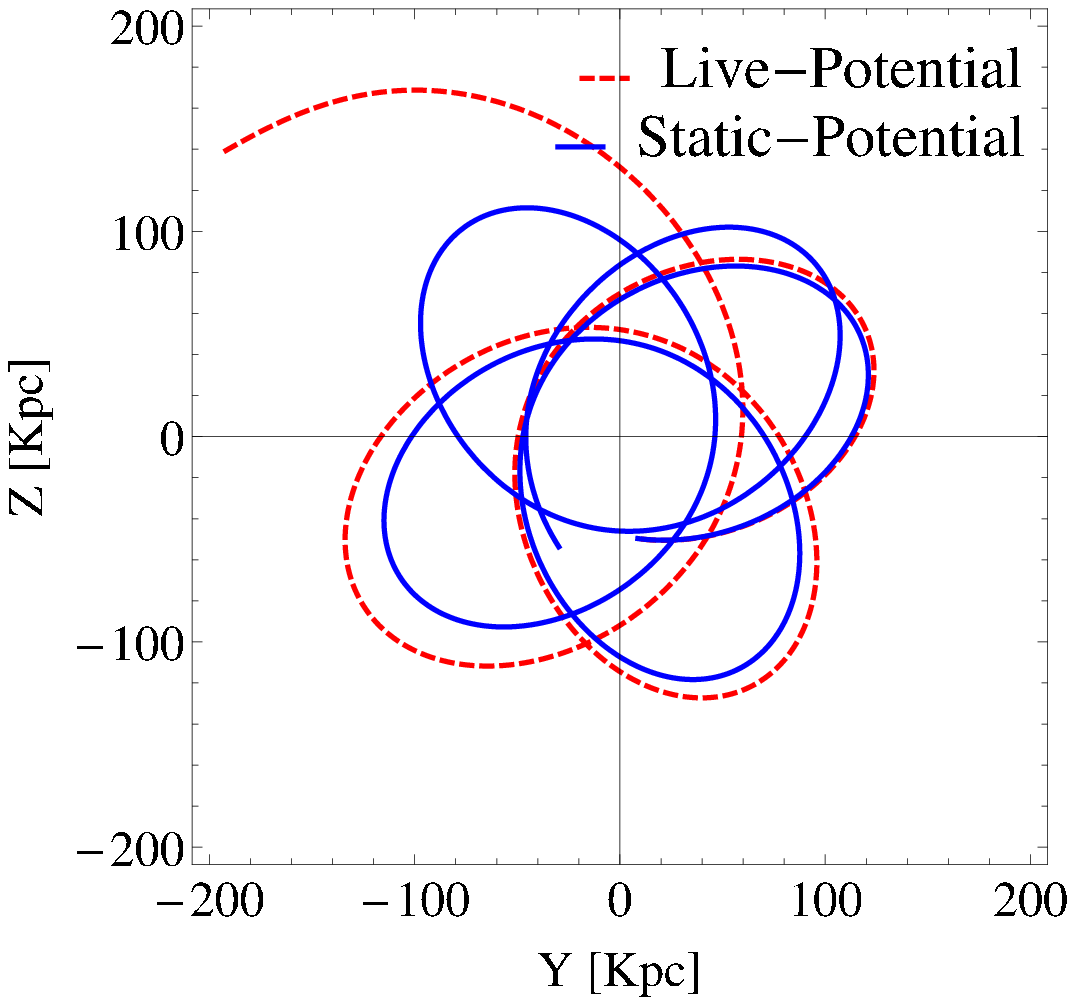}
    \includegraphics[width=80mm]{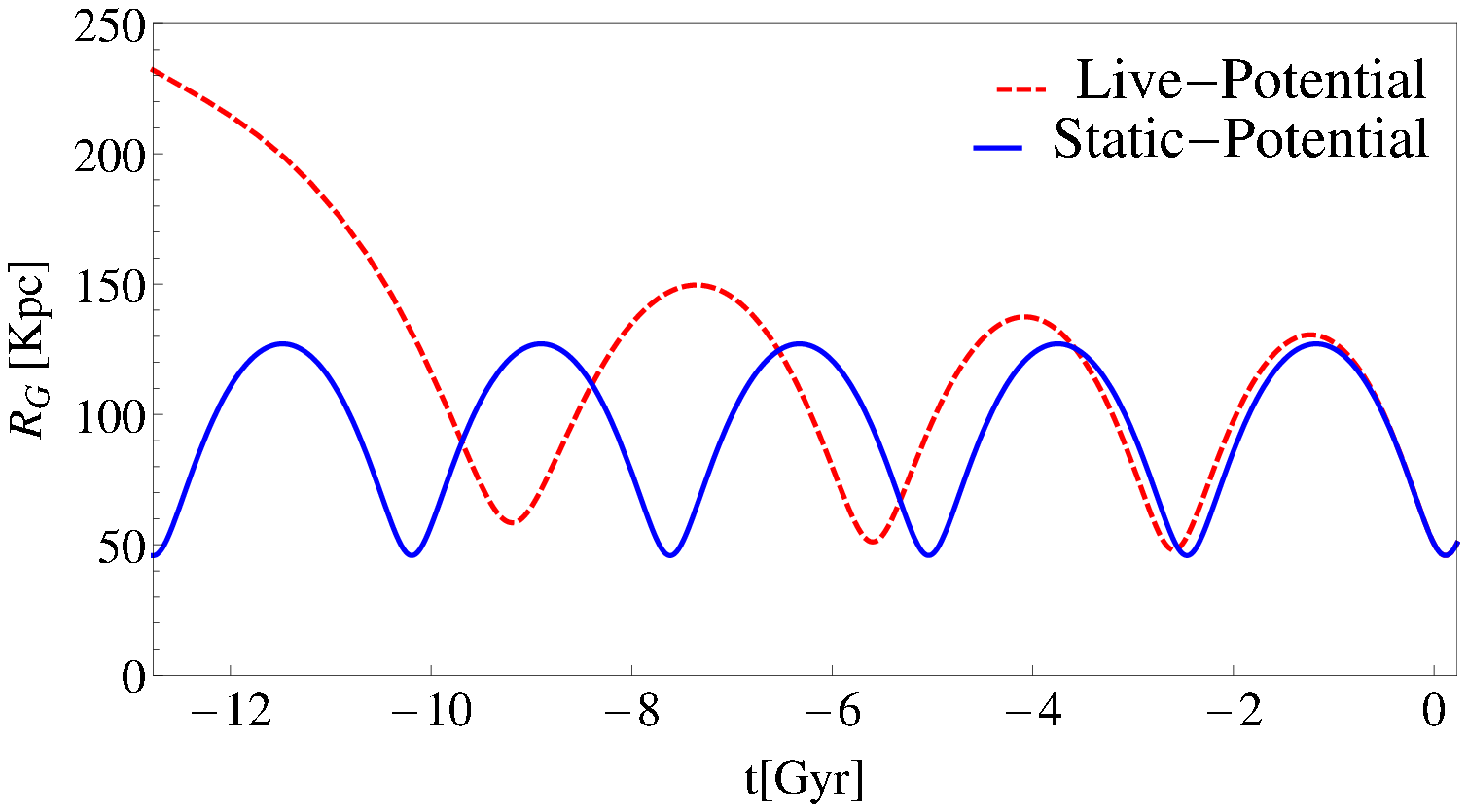}
	\caption{Top: section in the $Z-Y$ plane of the integrated trajectories of the LMC for the static potential (blue line) and the live potential (red dashed line).  The time span of the integration is $-13$ Gyr $\leq t \leq 0$. The initial conditions for position and velocity are taken from \citep{M05}. Bottom: the evolution of the Galactocentric distance of LMC as a function of time.  The red dashed line: time-dependent Galactic potential. Blue line:  static potential. The galactocentric distance of LMC, 13\,Gyr ago, would have been at $R_G=60$ kpc for Galactic model with static potential, while in the time-dependent Galactic potential it would have been at larger Galactocentric distance of about  $R_G=230$ kpc. }
		\label{fig:ZY_M05}
\end{figure}

Applying the same method for the LMC to numerically integrate the equation of motion, we extract its trajectory by backtracking orbit from its current position and velocity for 13 Gyr. Fig. \ref{fig:ZY_M05} depicts the planar shape of the orbits of the LMC in $Y-Z$ plane (upper panel).  We also show the evolution of the distance to the MW of the LMC in both static and live Galactic potentials in the lower panel of Fig. \ref{fig:ZY_M05}.
%We found that the current proper motion of the LMC would imply a period of about $~ 2.7$ Gyr in the static potential model, and $~ 4$ Gyr for a time-dependent Galactic potential model.

Another issue which should, in principle, take into account is the number of disc (i.e., the MW's disc) passage in both Galactic potentials. In fact, crossing  the disc would imply a strong orbital perturbation of the LMC and perhaps a gas shock which can lead to the formation of stars. We found that the number of disc passage with Galactocentric distances smaller than 100 kpc, in the static potential is five, while it is three in the live potential.

\begin{table}
	% title of Table
	\centering % used for centering table
	\caption{The coordinate, in kpc, and velocity component, in km s$^{-1}$, of LMC used in this work in a Galactocentric rest frame with the $z$-axis pointing towards the North Galactic Pole (NGP), the $x$-axis pointing in the direction from the Sun to the Galactic centre, and the positive $y$-axis is directed towards the Sun's Galactic rotation \citep{M05}.}
	\begin{tabular}{c c c } % centered columns (4 columns)
		\hline\hline %inserts double horizontal lines
		 $r(x,y,z)$ & 3D $v(x,y,z)$   \\ [0.5ex] % inserts table
		%heading
		\hline % inserts single horizontal line
		 (0,--43.9,--25.04)      & (--4.3,--182.45,169.8)\\
		\hline %inserts single line
	\end{tabular}
\label{table:LMC_orbital_param}
\end{table}

%\begin{figure}
%	
%	\includegraphics[width=90mm]{LMC-ZR}
%    	\caption{Orbit of LMC in Z-R plane where $R^2=X^{2}+Y^{2}$. The red dashed line is concerns the time-dependent potential while the blue line is for static potential.}
%	\label{fig:ZR_M05}
%\end{figure}

\section{Dynamical evolution of star clusters}  \label{dynamic}

\begin{figure*}
\begin{center}
\includegraphics[width=85mm]{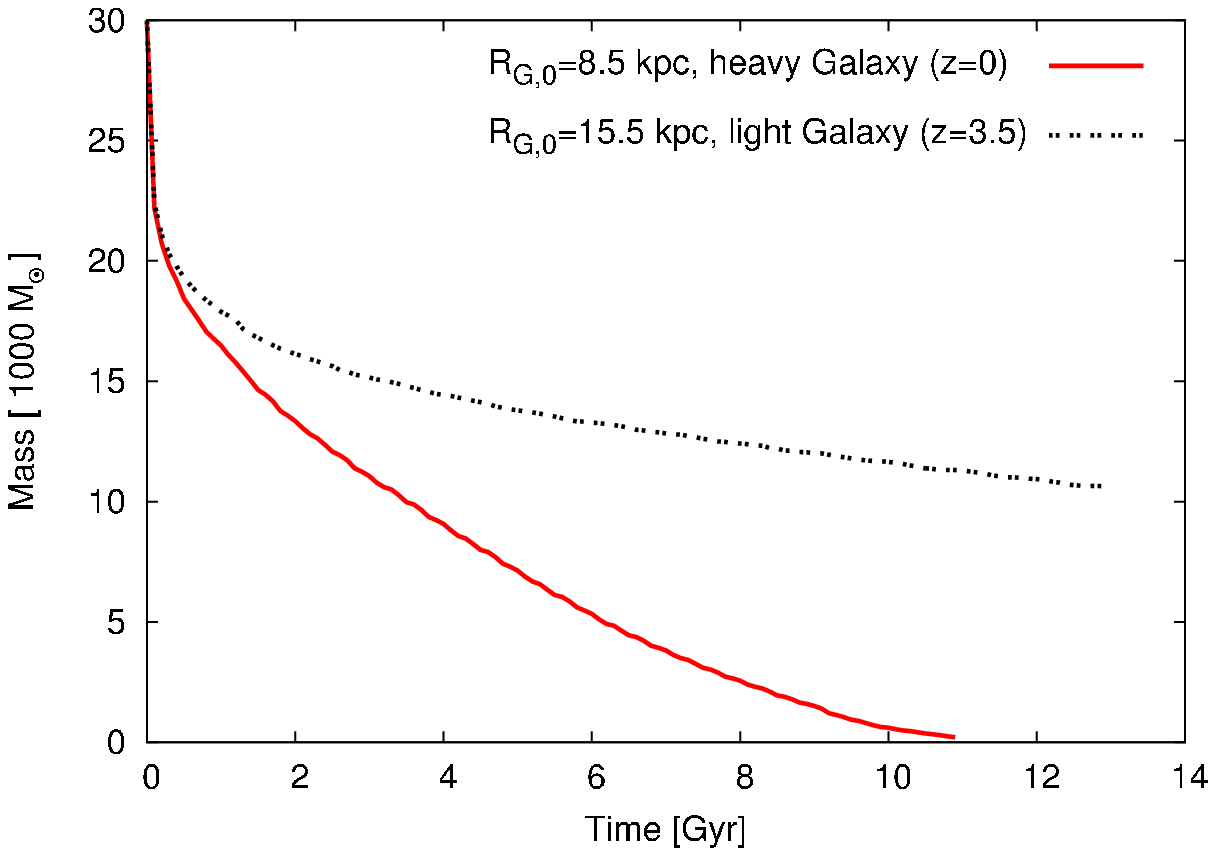}
\includegraphics[width=85mm]{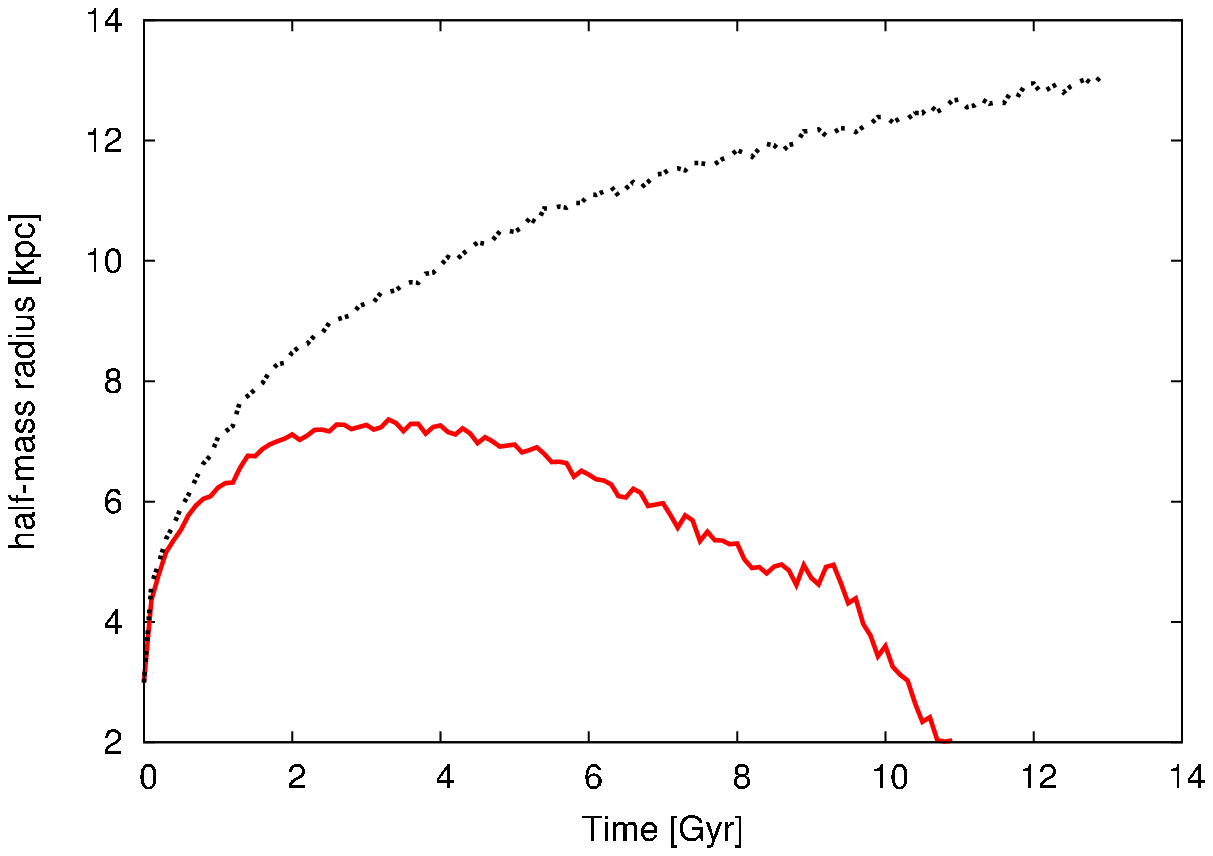}
\caption{The evolution of total mass (left-hand panel) and half-mass radius (right-hand panel) with time for two simulated star clusters orbiting within different Galactic potentials with parameters listed in Tables \ref{table:parameters} and \ref{table:parameters-12Gyr}.  Cluster moving in a circular orbit with a Galactocentric radius of $R_G=15.5$ kpc and affected by the gravitational field of our Galaxy with parameters at $z=3.5$ (corresponding to $T=-12$ Gyr) is shown by a black dashed line, and with Galactocentric radius of $R_G=8.5$ kpc  and affected by the gravitational field of our galaxy with parameters at $z=0$ (corresponding to the present day) is shown by a red solid line. For the cluster close to the galactic centre, expansion is limited by the strong tidal field and dissolves before a Hubble time.}
\label{massloss}
\end{center}
\end{figure*}

Our Galaxy hosts around 160 GCs. It is well understood that the gravitational potential of host galaxy has a direct influence on the survival and the evolution of GCs: they lose stars through tidal stripping and disc shocking. As already have shown by \cite{Praagman10}, varying the mass and concentration of the halo affects the rate at which the star cluster loses mass. Using $N$-body models of low-number star clusters, they found that increasing the halo mass and concentration drives enhanced mass-loss rates and, in principle, implies shorter dissolution time-scales.

Several studies have been addressed the evolution of clusters in time-dependent galactic potential by arbitrary switching tidal effects to mimic the accretion of a dwarf satellite on to a massive host galaxy \citep{Miholics14, Bianchini15},  or by rapidly varying evolutions (e.g., galaxy interactions and mergers) combining the galaxy simulations to star cluster simulations \citep{Renaud11, Renaud13, Renaud15}, or by evolving  star clusters in a cosmological environment \citep{Rieder13}.

This issue has been simplified by several authors. For example, \cite{Madrid14} have studied the impact of the host disc mass and geometry on the survival of star clusters by means of $N$-body simulations. They showed that a more massive disc enhances the mass-loss rate of an orbiting star cluster owing to a stronger tidal field such that doubling the mass of the disc halves the dissolution time of a star cluster located at $R_G=6$ kpc from the centre of Galaxy. They placed several of these simulations together, each time increasing the galaxy's mass, to represent a realistic mass growth history of the MW driven by mergers of satellite galaxies.

As we have shown by backward tracking of test objects, in the frame work of time-varying Galactic potential, the Galactocentric distance of a test object (like e.g., a star cluster in the MW) at $t=-12$\,Gyr is quite a bit larger than initial values at $t=0$, by a factor of about $~2$. In other words, in order to be a star cluster in its current position and velocity, it should be located at a larger Galactocentric distance in the past, i.e., 12\,Gyr ago, when the Galaxy was also much lighter than its present-day mass.  This implies that the cluster presumably was tidally underfilling \footnote{with $r_h/r_t$ values smaller than 0.05, where $r_h$ and $r_t$ are the cluster half-mass and tidal radii, respectively.} at the beginning of its evolution. Therefore, the slower  mass-loss rate of clusters initially lying inside their tidal radii, takes a longer time to lose a given amount of mass in comparison to tidally filling clusters (see e.g., \citealt{Baumgardt08, Marks08}). Indeed, in tidally limited clusters, the early evolution of massive stars leads to a rapid expansion, and hence a larger flow of mass over the tidal boundary. It may therefore help to dissolve them more rapidly. Tidally underfilling clusters, however, can survive this early expansion.

We therefore expect that different scenarios for Galactic potential (i.e.,  the time-varying versus the static invariant Galactic potentials), in principle, can lead to different evolution and survival of star clusters and consequently different depletion rates of satellite star clusters. Here, in this section we assess this difference by direct $N$-body simulations of GCs in a realistic MW-like potential using the code \textsc{nbody6}.

The current version of \textsc{nbody6} does not allow for a treatment of tides with an explicit time-dependent background potential. Within the current framework of \textsc{nbody6} and  in order to estimate the fate of a star cluster which its host galaxy grows with time we have calculated two independent models of star clusters with different masses and sizes for host galaxy.
\begin{itemize}
\item First we assume a star cluster in a circular orbit with a Galactocentric distance of $R_G=8.5$ kpc that evolves a \emph{"heavy"} galaxy with a mass and  geometrical parameters of the present-day MW-like potential listed in Table 1.
\item We then simulate the same star cluster in a circular orbit with Galactocentric distance of $R_G=15.5$ kpc within a \emph{"light"} host galaxy with parameters of the MW-like potential at $T=-12$\,Gyr given in Table \ref{table:parameters-12Gyr}. The reason for this choice of Galactocentric distance is that a cluster which is currently located at 8.5 kpc from the centre of the Galaxy had previously started to evolve (12\,Gyr ago) at a Galactocentric distance of $R_G=15.5$ kpc within the time-varying Galactic potential (See Fig. \ref{fig:YX_Sun}).
\end{itemize}

Our simulations include mass-loss driven by stellar evolution by using the \textsc{SSE/BSE} routines and analytical fitting functions developed by ~\citet{Hurley00} and  ~\citet{Hurley02}, the two-body relaxation, and a realistic treatment of the external tidal field. Both star clusters are evolving  with initial particle number of $N=50000$ (corresponding to $M\approx30000\msun$) that were distributed as a Plummer density profile (Plummer 1911) with the same initial half-mass radius of $r_{h,0}=$ 3 pc. The models  started with a Kroupa stellar IMF \citep{kroupa01, kroupa13}, which consists of two power laws with slope $\alpha_1=1.3$ for stars with masses between 0.08 and $0.5\msun$ and slope $\alpha_2=2.3$ for more massive stars. The range of stellar masses was chosen to be from 0.08 to 100$\msun$. The simulated clusters evolve on circular orbits at different galactocentric distances in the disc plane (i.e., the inclination angle of the orbits with respect to the galactic disc is zero). Our main focus is to study how quantities which can be checked observationally, as e.g., the slope of the MF or the size scale of the cluster change with time.

\subsection{The evolution of mass and characteristic radii}

\begin{figure}
\begin{center}
\includegraphics[width=85mm]{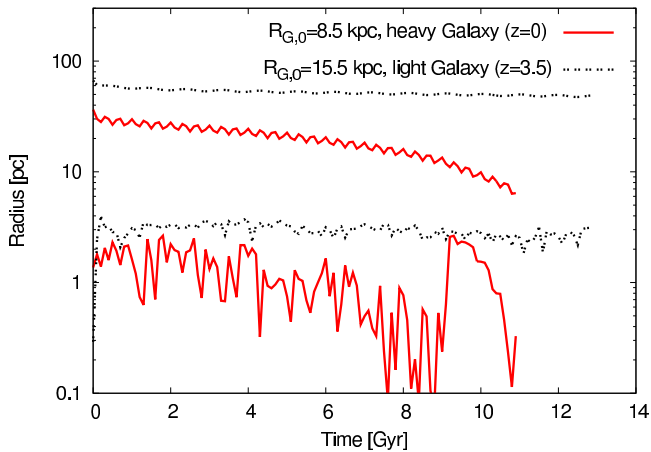}
\caption{The evolution of the core (lower curves) and tidal radius (upper curves) is plotted as a function of time in logarithmic scale.  Models are the same as Fig. \ref{massloss}. The tidal radius of cluster with orbit at $R_G=8.5$ kpc from the galactic centre decreases continuously due to ongoing mass-loss. Core collapse is reached at $T\simeq 9$\,Gyr for this cluster. But, the star cluster that evolves in a \emph{"light"} galaxy on a circular orbit with radius of $R_G=15.5$ kpc does not exhibit core collapse before a Hubble time.   }
\label{rcrt}
\end{center}
\end{figure}

\begin{figure}
\begin{center}
\includegraphics[width=85mm]{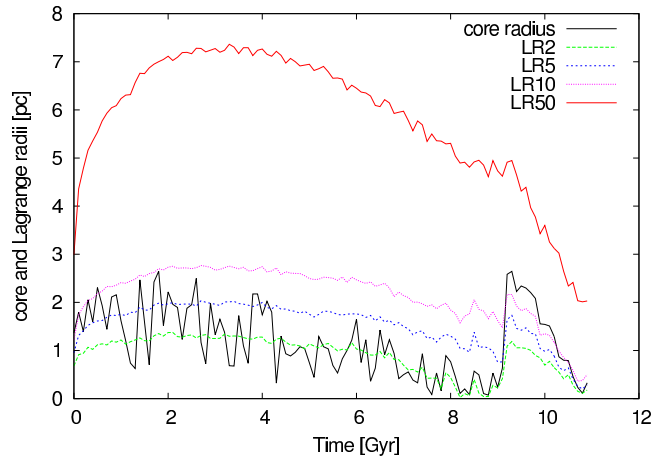}
\caption{Shown are the evolution of Lagrange radii and core radius of a cluster that evolves in a \emph{"heavy"} galaxy with orbit at $R_G=8.5$ kpc from the galactic centre as a function of time.  LR2, LR5, LR10, and LR50 are, respectively, the 2, 5, 10 and 50 per cent Lagrange radii.  Both the half-mass and core radii show an initial increase corresponding to stellar evolution and mass-loss from massive stars. Cluster reaches the end of initial core-collapse phase at $T\simeq 9$\,Gyr. The sharp change in the behaviour of the core at $T\simeq9$\,Gyr, when the size of the core suddenly increased might be linked to an interaction within the core involving a dynamically formed binary. }
\label{lag.rad}
\end{center}
\end{figure}

The evolution of the total mass of the simulated clusters with time for two different Galactic parameters as given in Tables \ref{table:parameters} and \ref{table:parameters-12Gyr} are plotted in Fig. \ref{massloss}. In fact, the long-term mass-loss for these clusters can be regarded as a runaway overflow over the tidal boundary. It can be seen, a star cluster that evolves in a \emph{"light"} galaxy on a circular orbit with radius of $R_G=15.5$ kpc has remaining mass of $10^4 \msun$, i.e., 30\% of its initial mass after a Hubble time of evolution, while a star cluster evolving in  a \emph{"heavy"} galaxy with $R_G=8.5$ kpc does not last for a Hubble time and dissolved after about 11 Gyr. %This is because, the former cluster that is tidally limited cannot reach a significantly larger half-mass radius.

Fig. \ref{massloss} (right-hand panel) depicts the time-evolution of the 3D half-mass radius of simulated star clusters at different galactocentric distances. Clusters start with an initial half-mass radius of 3 pc and undergo an expansion triggered by stellar evolution within the
first Gyr.  The half-mass radius of a simulated star cluster evolving within a "\emph{heavy}" Galaxy (with the present-day values of parameters of MW listed in Table \ref{table:parameters}) at 8.5 kpc from the galactic centre, reaches a maximum value, which appears to be clearly linked to its galactocentric distance,  before it decreases again until the cluster dissolves after 11 Gyr of evolution. As shown in Fig. \ref{massloss}, the evolution of a star cluster orbiting at $R_G=$15.5 kpc under the tidal field of a "\emph{light}" Galaxy with the parameters of the MW-like potential at $T=-12$\,Gyr (Table \ref{table:parameters}) appears to have not reached its tidal limit yet and keeps expanding till the end of the simulation. This is easy to understand: a smaller orbital radius leads to a faster disruption and a smaller half-mass radius after 13 Gyr evolution owing to the enhanced mass-loss driven by the galactic tide, and the stronger cut-off it inflicts on the clusters.

Another useful diagnostic of the difference between models at different galactocentric distances is provided by the core \footnote{In our simulations, the core radius is a density weighted average distance of each star to the point of highest stellar density within the cluster \citep{Casertao85, Aarseth03}, while in observational studies the core radius is generally defined as the radius where the surface brightness falls to half its central value (King 1962).} and tidal radii. Their time evolutions are illustrated in Fig. \ref{rcrt}.  Accentuated mass-loss at $R_G=8.5$ kpc as seen in Fig. \ref{massloss} precipitates the onset of core-collapse. The time of core-collapse is usually determined by the moment of core bounce, which is seen in the time evolution of the core radius or density radius \citep{Fujii14}.  However for some models it is difficult to distinguish the core-collapse, because there does not seem to be a peak in the density evolution or a depression in the core radius. The core radius shows a minimum at $\sim 9$\,Gyr for a star cluster evolves at $R_G=8.5$ kpc from the centre of "\emph{heavy}" galaxy model which we identify as the moment of the initial core-collapse phase ends (Fig. \ref{rcrt}). According to Fig. \ref{lag.rad}, the core radius expands in the beginning (within the first 2\,Gyr) due to the weak tidal field, but eventually at $T\simeq9$\,Gyr it goes into a small core collapse. But since it is also close to being disrupted, the core collapse does not look like the usual deep contraction. The cluster expands in a jump, probably due to the ejection of the central binary or even a binary--binary ejection. This leaves the core much less bound than before the ejection, and hence the cluster expands and dissolves in an instant.  The evolution of Lagrange radii also supports this conclusion as they go up at $T\simeq9$\,Gyr, and one can conclude this is core collapse.  Short-term effects on the evolution of core radius are often related to the presence of high-energetic binaries, probably comprised of two black holes in the core.

The cluster in orbit of 15.5 kpc ("\emph{light}" Galaxy) does not reach the end of the core-collapse phase in 13\,Gyr.   Therefore, we conclude that the core radius is affected by changes in tidal forces on the cluster, in agreement with \cite{Madrid12} who have found that the galactocentric distance of a star cluster has an impact on its core radius and the onset of core collapse. This is in contradiction with the conclusion reached by \cite{Miholics14} who showed that the core radius of a cluster will depend on the initial structural conditions of the cluster and will not be affected by its tidal history. It can be seen that the tidal radius of the cluster in the heavier (present day) galaxy is smaller (tidally limited) than the cluster in the lighter (12\,Gyr ago) galaxy. We therefore conclude that clusters survive longer in an evolving galaxy than in a galaxy which is kept static.
%Core collapse is reached at $\sim 9$ Gyr for a star cluster evolves at $R_G=8.5$ kpc from the center of heavy galaxy model  (Fujii & Portegies Zwart, 2013)

As expected, galaxy with masses between these two extremes (i.e., a time-dependent galaxy mass model) define intermediate regime of mass-loss. One can therefore conclude that after a Hubble time of evolution,  a star cluster has less mass in a static potential than a simulated cluster evolving in a live potential. This is because, in the static potential the galaxy mass on average (over time) is larger than the live potential. As already have shown by several authors, increasing the mass of host galaxy accelerate the destruction time of star clusters \citep{Praagman10, Madrid14}.

However, the evolution of a star cluster strongly depends on its filling factor ($r_h/r_t$). The present-day underfilling clusters (i.e., $r_h/r_t < 0.05$) would remain underfilling in an evolving-potential over the whole 12\,Gyr of evolution. This is because of the larger $R_G$ and lighter $M_G$ compared to a static potential (see Fig. \ref{rcrt}).  So one would expect the cluster's final size to be nearly similar to that of the static case. But, a present-day tidally limited cluster would be underfilling for a while in an evolving galactic potential.

\subsection{The evolution of the MF}

\begin{figure}
\begin{center}
\includegraphics[width=85mm]{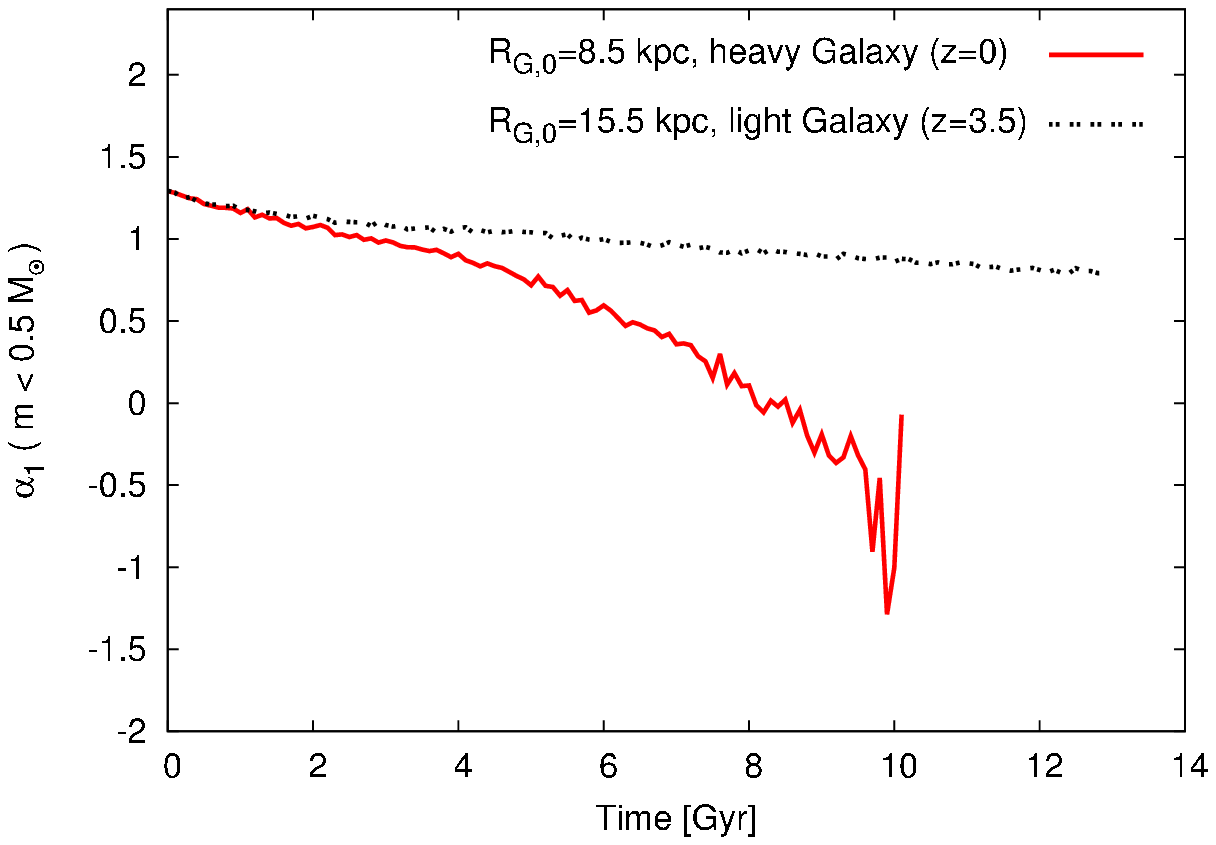}
\includegraphics[width=85mm]{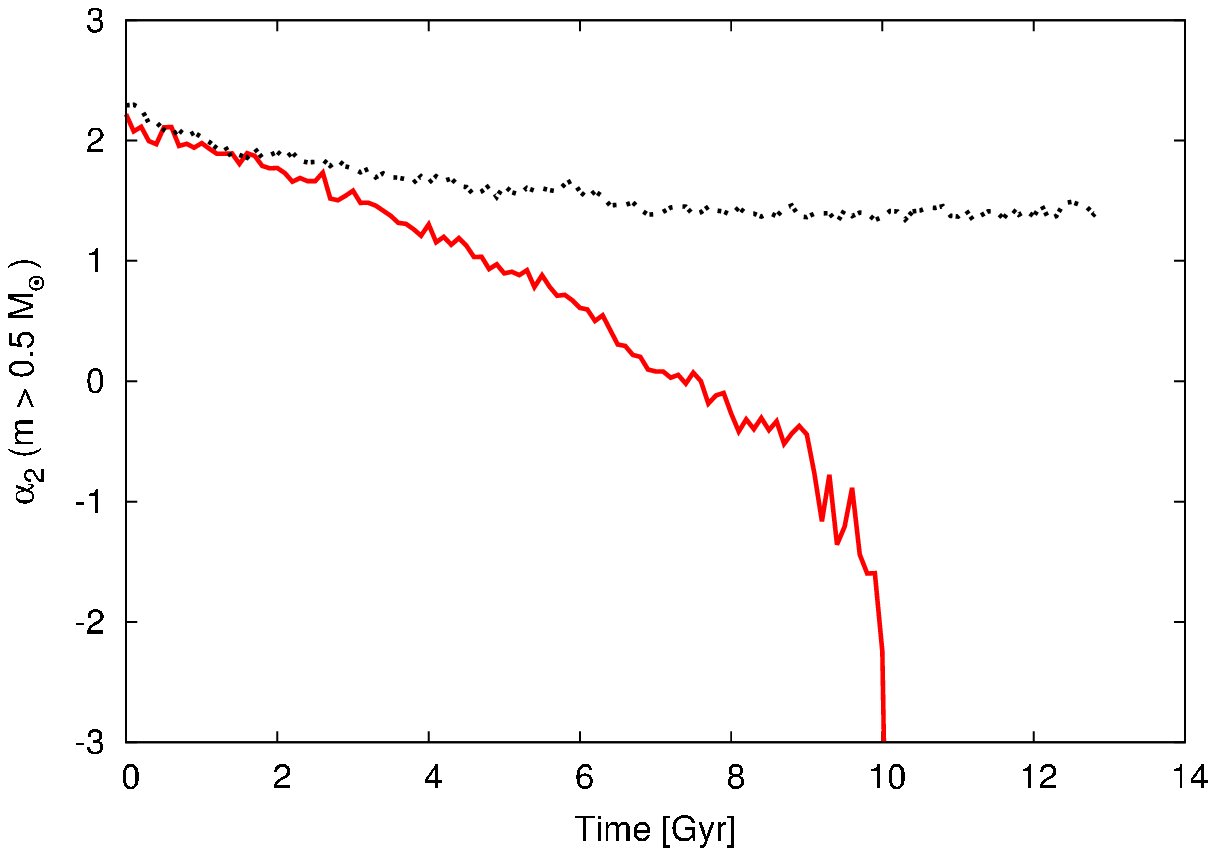}
\caption{The evolution of the global stellar MF-slope for low-mass ($m\leq 0.5 M_\odot$, top panel) and high-mass stars ($m\geq 0.5 M_\odot$, bottom panel) within the half-mass radius is plotted as a function of time.  Models are the same as Fig. \ref{massloss}. A flatter MF is the direct result of increased tidal stripping of outer region stars that are preferentially low in mass due to dynamical mass segregation.}
\label{alpha}
\end{center}
\end{figure}

The stellar MF is one of the most important observable parameters that changes through the dynamical evolution of star clusters. It is evident that the two-body relaxation driven evaporation through the tidal boundary of a star cluster gives rise to a significant correlation between the MF-slope and the strength of tidal field of host galaxy. Because of the dynamical mass segregation occurs due to two-body relaxation, the evaporation rate  is larger for low-mass stars than it is for high-mass ones (Giersz \& Heggie 1997, Baumgardt \& Makino 2003). Thus, the preferential escape of low-mass stars leads to the flattening of the MF as the dynamical evolution of a star cluster proceeds. It is shown that the evolution of MF-slope is faster for clusters at smaller galactocentric distances (Vesperini \& Heggie 1997), i.e., experience a stronger tidal field (Webb et al. 2013).

The canonical IMF as observed in young star clusters in the MW is often expressed as a two-part power-law function ($\frac{dN}{dm}\propto m^{-\alpha}$) with near Salpeter-like slope above $0.5\,\mbox{M}{\odot}$ (i.e., $\alpha_2=2.3$; \citealt{Salpeter55}), and a shallower slope  of $\alpha_1=1.3$ for stars
in the mass range $0.08-0.5\,\mbox{M}{\odot}$ \citep{kroupa01, kroupa13}.

Fig. \ref{alpha} depicts the evolution of the MF-slope at the low-mass end as a function of time.  The slope was determined from a fit to the distribution of stars with masses $m \leq 0.5$. In both cases the MF flattens as the cluster loses stars. Therefore, as the galactocentric distance of a GC decreases, the strength of the stellar mass-loss driven by two-body relaxation increases, and hence the amount of the flattening of the MF enhances.  Fig. \ref{alpha} confirms that the slope of the MF in a cluster orbiting at $R_G=8.5$ kpc within a \emph{"heavy"} Galaxy  changes significantly as compared to a cluster evolving in a circular orbit with a Galactocentric radius of $R_G=15.5$ kpc within a \emph{"light"} galaxy model.

Recent observational work on a number of MW GCs have shown that the global MF-slope in the low-mass range is significantly shallower than a canonical MF-slope of about 2.3  (Kroupa 2001, see e.g.  \citealt{De Marchi07,Jordi09,Paust10,Frank12,Hamren13}).
However, the preferential loss of low-mass stars due to two-body relaxation would be a natural explanation for the observed MF depletion \citep{Baumgardt03},  for diffuse outer halo clusters such as Pal~4 and Pal~14  (i.e., a low mass together with a large half-mass radius),  the present-day two-body relaxation time is of the order of a Hubble time. Therefore, relaxation should be inefficient in these clusters and the observations should be an indication for primordial mass segregation \citep{Zonoozi11, Zonoozi14}. Our findings in this section show that considering a live potential for our Galaxy makes it even more difficult to explain the observed MF flattening. This is because, the changes in the MF due to the tidal stripping is less in a live potential compared to the static potential model.

\section{Conclusions}\label{Sec:Conclusions}

Many authors usually use the static potential for our Galaxy, as the common assumption is that it remains unchanged during the orbital integration.
In this paper, we have investigated the influence of the time-dependence of the Galactic potential on the orbital history of the halo objects and its consequences on their internal evolution. First, we numerically integrated backwards the orbits of different test objects over a Hubble time, located in different Galactocentric distances within both static and live (cosmologically motivated) Galactic potentials to assess the possible differences.

It turns out that, the static and live potential do yield different trajectories for our test objects orbiting in different Galactocentric distances. We have shown that the spatial extinction of the orbit's section in the coordinate planes is larger for live potential w.r.t static potential, such that in a live potential, the birth of the objects, 13\,Gyr ago, would have occurred at significantly larger Galactocentric distances, compared to the objects orbiting in a static potential.

As a concrete example of motion within the Galactic halo, we also used the backward motion of the LMC in both static and live Galactic potentials. In addition to the different trajectories of LMC we uncovered here, we found that the orbital period of the LMC around the MW is about $~ 2.7$ Gyr in the static potential model, while it is $~ 4$\,Gyr in a time-dependent Galactic potential model. This is important because it is believed that the kinematics and morphology of the MS follows the orbits of the LMC. We furthermore found that, in the static potential, the number of disc passage with Galactocentric distances smaller than 100 kpc is five, while it is three in the live potential.

We finally investigated the impact of the assuming a live potential on the dynamical evolution of star clusters by means of the collisional $N$-body code, \textsc{nbody6}.  Since the current version of \textsc{Nbody6} does not allow for a treatment of live potential of host galaxy with an explicit time-dependence,  we calculated two models of star cluster with different masses and sizes for host galaxy (which represents the galaxy at present and 12\,Gyr ago),  to roughly estimate how a cluster's half-mass radius, the total mass, and the MF-slope develops over the time.

We followed the evolution of clusters at different Galactocentric distances with different Galaxy mass models, and found that the weaker mass-loss of clusters evolving in a weaker tidal field (i.e., at a larger Galactocentric distance and within a light-mass Galaxy) leads to a significantly larger final size.

Our computations demonstrate that for two star clusters moving in circular orbits with different Galactocentric distances and within different Galactic models, one with $R_G=8.5$ kpc and within a heavy-mass Galaxy (with the present-day parameters of our Galaxy) and one with $R_G=15.5$ kpc evolving in a light-mass Galaxy (with parameter values at 12\,Gyr ago), the star cluster with smaller Galactocentric distance will have:

\begin{itemize}
\item a stronger tidal truncation and a smaller size
\item an enhanced mass-loss rate and a shorter dissolution time
\item a flatter MF
\end{itemize}

Since a galaxy with parameters between these two extremes defines intermediate regime of mass-loss, one can therefore conclude that over a Hubble time of evolution in a semicosmological time-dependent Galaxy model, the star cluster has more mass than a simulated cluster evolving in a static invariant potential; this is because in the static potential the galaxy mass on average (over time) is larger than the live potential. This implies that assuming a static potential for our Galaxy (as it is often done) leads to an enhancement of mass-loss rate,  an overestimation of the dissolution rates of GCs, and  an underestimation of the final size of star clusters.

Consequently, after a Hubble time of evolution in the framework of a live Galactic potential, we expect to see the more survival of star clusters as compared to simulated star clusters, which evolve in a galaxy with a constant mass components. Clearly, we do not claim that the exercise above represents a realistic effect of time-dependence of Galactic potential on the evolution of star clusters.  Investigating the fate of a star cluster within a galaxy which grows with time and comparing with the different $N$-body methods (e.g. \textsc{amuse}; \citealt{Rieder13} and \textsc{nbody6tt}; \citealt{Renaud11}) is our upcoming project (Zonoozi et al., in preparation).

\section*{Acknowledgements}

This paper is dedicated to Professor Yousef Sobouti, the founder of IASBS,  for his tireless and distinguished efforts in promoting the scientific research in the theoretical astrophysics in Iran. We would like to thank the referee for constructive comments and suggestions. HH  would like to thank Andreas K\"{u}pper and Holger Baumgardt for helpful comments.  This work was made possible by the facilities of Graphics Processing Units at the IASBS.

%We would like to thank the referee for constructive comments and suggestions. AHWK acknowledges support through Hubble Fellowship grant HST-HF-51323.01-A awarded by the Space Telescope Science Institute, which is operated by the Association of Universities for Research in Astronomy, Inc., for NASA, under contract NAS 5-26555. This work was made possible by the facilities of Graphics Processing Units at the Institute for Advanced Studies in Basic Sciences (IASBS).

\bsp \label{lastpage}

\end{document}